%% file: anatomically_informed_arxiv.tex
\documentclass{article}
\usepackage{spconf,amsmath,graphicx}
\DeclareMathOperator*{\argmin}{argmin}
\usepackage{amssymb}

\usepackage{setspace}
\usepackage{amsmath}
\makeatletter
\g@addto@macro\normalsize{%
  \setlength\abovedisplayskip{3pt}
  \setlength\belowdisplayskip{3pt}
  \setlength\abovedisplayshortskip{3pt}
  \setlength\belowdisplayshortskip{3pt}
}

\title{Anatomically-Informed Multiple Linear Assignment Problems for
  White Matter Bundle Segmentation}
\name{\normalsize{Giulia Bert\`o$^{1,2}$, Paolo Avesani$^{1,2}$, Franco Pestilli$^{3}$, Daniel Bullock$^{3}$, Bradley Caron$^{3}$, Emanuele Olivetti$^{1,2}$
}   
\thanks{This work was partially supported by NSF IIS-1636893, NSF BCS-1734853, NIH NIMH ULTTR001108, a Microsoft Research Award, a Google Cloud Award, the Indiana University Areas of Emergent Research initiative “Learning: Brains, Machines, Children,” and Indiana University Pervasive Technology Institute to F.P. 
}
}
\address{
   $^{1}$ \small{NeuroInformatics Laboratory (NILab),
     Bruno Kessler Foundation, Trento, Italy} \\ 
   $^{2}$ \small{Center for Mind and Brain Sciences (CIMeC), University of Trento, Italy} \\
   $^{3}$ \small{Indiana University, Bloomington IN 	  
    47405, USA}}
 
\begin{document}
%
\maketitle
\begin{abstract}
Segmenting white matter bundles from human tractograms 
is a task of interest for several  applications. 
Current methods for bundle segmentation consider either only prior knowledge about the relative anatomical position of a bundle, or only its geometrical properties. 
Our aim is to improve the results of segmentation by proposing a method that takes into account information about both the underlying anatomy and the geometry of bundles at the same time. 
To achieve this goal, we extend a state-of-the-art example-based method based on the Linear Assignment Problem (LAP) by including prior anatomical information within the optimization process.
The proposed method shows a significant improvement with respect to the original method, in particular on small bundles.

\end{abstract}

\input{introduction}

\input{methods}

\input{experiments}

\input{discussion}

\bibliographystyle{IEEEbib}
\bibliography{bib-anatomical}

\end{document}

%% file: introduction.tex
\section{Introduction}
\label{sec:introduction}


Segmenting anatomical structures in the white matter of the human
brain is useful in many different applications, such as surgical
planning, population studies, and diagnosis or monitoring of brain
diseases \cite{odonnell2007automatic, yeatman2012tract}.

The information about the orientation of the fibers composing such
anatomical structures can be estimated in-vivo by diffusion Magnetic
Resonance Imaging (dMRI) techniques. By means of tractography, the
paths of hundreds of thousands of fibers composing the white matter
can be mathematically represented by 3D polylines called
\textit{streamlines}. White matter bundle segmentation aims to
virtually group together streamlines that have an analogous shape
and pass through the same anatomical brain regions
into anatomically meaningful structures, known as \textit{bundles},
e.g. the uncinate fasciculus (UF) (see Figure \ref{fig:example}).

To overcome the limitations of manual segmentation \cite{catani2002virtual, wakana2007reproducibility}, which is very time
consuming, in recent decades several
automatic methods have been developed. They can be divided into
two categories  \cite{garyfallidis2017recognition}: (i) connectivity-based, 
and (ii) streamline-based methods.  Connectivity-based
approaches aim to extract white matter bundles by means of predefined
brain Region of Interest (ROIs) that the streamlines are supposed to pass (or not-pass) through \cite{yeatman2012tract, wassermann2016white}. 
Streamline-based techniques are able to
segment white matter bundles of interest by grouping together
streamlines according to their geometrical similarity (clustering-based)
\cite{odonnell2007automatic, zhang2008identifying} or by exploiting 
the geometric information from previously segmented bundles, usually 
validated by experts, that are used as examples (example-based)
\cite{yoo2015example, garyfallidis2017recognition, sharmin2018white}. Example-based
methods have shown to outperform connectivity-based
methods because these last ones strongly depend on the registration
between subject and atlas and on the quality of 
the parcellation \cite{garyfallidis2017recognition,
  sharmin2018white}. Another disadvantage of connectivity-based
techniques is that they do not take into account the shape of the
streamlines, but only anatomical regions. On the
other hand, clustering-based and example-based methods are based only
on geometrical properties of the streamlines without considering any
prior anatomical information about the bundle.

We aim to improve the results of bundle segmentation by considering
information about both the shape of the streamlines and about the relative anatomical position of the bundle of interest. In order to achieve this goal,
we propose to extend the example-based method proposed in
\cite{sharmin2018white}, whose implementation is publicly available,
by including additional anatomical information within the optimization
process of the Linear Assignment Problem (LAP)
\cite{bijsterbosch2010solving}. Specifically, the extra information is
given by taking into account the location of the endpoints of
the streamlines and the proximity of the streamlines to specific
anatomical ROIs predefined in the literature. We select
\cite{sharmin2018white} as a reference method since it is based on the
LAP and has shown to provide better results than those based on the
nearest-neighbor algorithm, such as
\cite{yoo2015example} and \cite{garyfallidis2017recognition}.

Our contributions are the following: (i) to extend the work in
\cite{sharmin2018white} by including anatomical information in
addition to the geometrical one, hence showing that it is possible to
combine the best of streamline-based and connectivity-based methods;
(ii) to show that small bundles are more difficult to be accurately
segmented than large bundles and that anatomical information helps in
reducing such difference.

We perform example-based bundle segmentation of 12 different bundles,
each of them in 30 different subjects, for a total of 360 segmented
bundles. We compare the proposed method with \cite{sharmin2018white}
by evaluating the results with the Dice Similarity Coefficient (DSC) to
measure the overlap between the segmentation and the ground truth. 
We support scientific reproducibility and openness by publishing our methods as brainlife.io/apps.

In Section \ref{sec:methods} we present the LAP method and the
proposed extension, Section \ref{sec:experiments} contains the
experiment design and the results, while Section \ref{sec:discussion}
presents a short discussion.


%% file: methods.tex
\section{Methods}
\label{sec:methods}


\subsection{Basic notation}
\label{sec:notation}

We denote a streamline with a sequence of $n$ points as
$s = ( \mathbf{x}_1, \ldots , \mathbf{x}_n)$, where
$ \mathbf{x}_i \in \mathbb{R}^3 $, $\forall i$. Usually, $n$ is in the
order of $10^1 - 10^2$ and differs across streamlines. The entire set
of streamlines of the white matter of a brain is known as the \textit{tractogram},
$T = \{s_1, \ldots, s_M\}$, where in general $M$ is in the order of
$10^5 - 10^6$. A white matter bundle $b\subset T$,
$b = \{s_1, \ldots, s_k\}$, is a group of streamlines with a specific
anatomical meaning, where $k \ll M$, and $k$ differs across bundles.

Several distance functions are available in the literature in order to
quantify the geometrical distance between two streamlines.  One of the
most common is the Mean of Closest distances (MC)
\cite{zhang2008identifying}:
$ d_{MC}(s_a, s_b) = \frac{d_m(s_a, s_b) + d_m(s_b, s_a)}{2} $ where
$d_m(s_a, s_b) = \frac{1}{|s_a|} \sum_{\mathbf{x}_i \in s_a}
\min_{\mathbf{x}_j \in s_b} \|\mathbf{x}_i - \mathbf{x}_j\|$
and $\| \cdot \|$ is the Euclidean distance.

\vspace{-0.2cm}

\subsection{The Linear Assignment Problem for Segmentation}
\label{sec:lap}
Given two sets of objects, $A = \{a_1, ... a_L\}$ and
$B = \{b_1, ... b_L\}$, and the cost matrix
$C =\{c_{ij}\}_{ij} \in \mathbb{R}^{L \times L} $, where $c_{ij}$ is
the cost of assigning $a^A_i$ to $b^B_j$, the Linear Assignment
Problem (LAP) \cite{bijsterbosch2010solving} aims to find the
one-to-one correspondence between the objects in A and the objects in
B by minimizing the total cost:
\begin{equation}
  \label{eq:RLAP}
  P^* = \argmin_{P \in \mathcal{P}} \sum_{i, j=1}^L c_{ij} p_{ij}
\end{equation}
where $P = \{p_{ij}\}_{ij} \in \mathcal{P}$ is a permutation matrix
and $P^*$ is the optimal assignment\footnote{If the size of the two
  sets of objects is different, i.e. $|A| \neq |B|$, the problem is
  called Rectangular Linear Assignment Problem (RLAP) and $P$ becomes
  a partial permutation matrix \cite{bijsterbosch2010solving}.}.
If multiple cost matrices need to be jointly optimized, they can 
be linearly combined.  One of the most efficient algorithm to solve the
LAP is the Jonker-Volgenant algorithm (LAPJV)
\cite{bijsterbosch2010solving}.

In the case of example-based bundle segmentation, in \cite{sharmin2018white}
they consider the two sets of objects as being (i) the example bundle of $A$,
$b^A = \{s_1^A, \ldots, s_k^A\}$, and (ii) the tractogram of a subject
$B$, $T^B = \{s_1^B, \ldots, s_M^B\}$, from which we aim to segment
the same anatomical bundle.  The goal of bundle segmentation is to
find the optimal correspondence of all the streamlines in $b^A$ with
those in $T^B$. In \cite{sharmin2018white}, the cost matrix is equal
to the distance matrix $D$ between the two sets of streamlines, in
which each element is given by $c_{ij}=d_{MC}(s_i^A, s_j^B)$.
Intuitively, the closer two streamlines are to each other, 
the more likely they belong to the same
anatomical structure. The segmentation from multiple examples 
proposed in \cite{sharmin2018white} is obtained by merging together 
the solution of multiple LAP solved individually. This is done through 
a refinement step that classifies the streamlines based on a majority rule.

\vspace{-0.2cm}

\subsection{Anatomically-Informed cost matrices}
\label{sec:anatomy_cost_matrices}
In order to include anatomical information into the LAP, we extend the
cost matrix $D$ by adding two weighted anatomically-informed cost matrices: the
\emph{endpoint-distance matrix} $E$ and the \emph{ROI-based distance
  matrix} $R$. Then, the new cost matrix becomes
$C = \lambda_D D + \lambda_E E + \lambda_R R$.

\vspace{-0.2cm}

\subsubsection{Endpoint-based distance matrix}
\label{sec:endpoint}
White matter fibers serve as connections between areas of the brain
at their terminations. For this reason, from an anatomical and
functional point of view, two streamlines 
with neighboring endpoints are assumed to play a similar role, regardless of their geometry.
Based on this idea, we
propose to build a new cost matrix $E$ by defining the
\emph{endpoint-based} distance, $d_{\text{END}}$, between two 
streamlines  $s_a$ and $s_b$ as the mean Euclidean distance  of their
corresponding endpoints: 
$ d_{\text{END}}(s_a, s_b) = \linebreak \frac{\min(\|\mathbf{x}_1^a -
  \mathbf{x}_1^b\|, \|\mathbf{x}_1^a - \mathbf{x}_{n_b}^b\|) +
  \min(\|\mathbf{x}_{n_a}^a - \mathbf{x}_1^b\|, \|\mathbf{x}_{n_a}^a -
  \mathbf{x}_{n_b}^b\|)}{2}$,
where $\{\mathbf{x}_1^a$,$\mathbf{x}_{n_a}^a\}$ are the endpoints of
$s_a$ and $\{\mathbf{x}_1^b$,$\mathbf{x}_{n_b}^b\}$ are those of
$s_b$.

\vspace{-0.2cm}

\subsubsection{ROI-based distance matrix}
\label{sec:roi}
Anatomically, bundles may be defined with respect to specific ROIs 
that they can cross or touch
\cite{catani2002virtual, wakana2007reproducibility, yeatman2012tract, wassermann2016white}.  Unfortunately,
such regions cannot be easily and precisely defined in the subject
space, but are often obtained by registration from an atlas, a step
that is intrinsically limited by the inherent differences between the
atlas and the specific subject. Nevertheless, such anatomical
information is of primary importance. To create an additional cost
matrix $R$ with such anatomical information, we first define the
distance between a streamline and a single ROI, as the minimum
Euclidean distance between them:
$d_{min}(s, \text{ROI}) = \min_{\mathbf{x}_i \in s, v_j \in
  \text{ROI}}\|\mathbf{x}_i - coord(v_j)\|$,
where $coord(v)$ are the 3D coordinates of the center of the voxel $v$
belonging to the ROI. Given a set of $N$ ROIs defining a bundle, the
distance between a streamline and them can be defined as the average
of the distances to each ROI. With such building block, we define the
ROI-based distance between two streamlines $s_a$ and $s_b$, for a
given set of ROIs as
$d_{\text{ROIs}} = | \frac{1}{N}\sum_{i=1}^N d_{\text{min}}(s_a,
\text{ROI}_i) - \frac{1}{N}\sum_{i=1}^N d_{\text{min}}(s_b,
\text{ROI}_i)|$,
meaning that two streamlines at similar distances from the ROIs are
more likely to belong to the same anatomical structure.


%% file: experiments.tex
\section{Experiments}
\label{sec:experiments}

\subsection{Materials}

We randomly selected 130 healthy subjects from the publicly available
Human Connectome Project (HCP) dMRI
dataset \cite{sotiropoulos2013advances} (90 gradients; b=2000; voxel
size=1.25 mm 
isotropic). For each subject, tractograms of 750k
streamlines were obtained using constrained
spherical deconvolution (CSD) \cite{tournier2007robust} and
ensemble probabilistic tracking \cite{takemura2016ensemble} 
(step size=0.625 mm, curvature parameters = 0.25, 0.5, 1, 2 and 4 mm).

Since example-based methods need accurate bundle segmentations to use both as example and to evaluate the results, we built a visually inspected ground truth dataset using a semi-automatic technique as follows. 
 First, from each of the 130 tractograms, we segmented the 20 major associative bundles using the Automated Fiber Quantification (AFQ) algorithm \cite{yeatman2012tract}. Then, in order to have consistent bundles across subjects, given a bundle, we identified those that do not deviate more than the $20\%$ from the median number of streamlines, obtaining on average 50 segmentations per bundle. Finally, we visually inspected each segmentation and filtered out the outliers in order to have 30 segmentations per bundle.
We then selected 12 bundles (6 left and 6 right) per subject, which we subdivided into two groups based on their number of streamlines\footnote{This is not an absolute definition of small and large bundles, but only a relative definition within the group of bundles considered in this work.}: 
the small bundles are Cingulum Cingulate (CGCl and CGCr), Cingulum Hippocampus (CGHl and CGHr) and Uncinate Fasciculus (UFl and UFr), while the large bundles are Thalamic Radiation (TRl and TRr), Corticospinal tract (CSTl and CSTr) and Arcuate Fasciculus (AFl and AFr).

To evaluate the results of the proposed method, we measured the degree of overlap between the estimated bundle 
$\hat{b}^B$ and the true bundle
$b^B$, using the Dice Similarity Coefficient (DSC)
at the voxel-level\footnote{The DSC takes values between 0 (no overlap) and 1 (perfect overlap).}:
  $DSC = 2 \frac{|v(\hat{b}^B) \cap
    v(b^B)|}{|v(\hat{b}^B)| +
    |v(b^B)|} $
where $v(b)$ is the set of voxels crossed by the streamlines of bundle
$b$ and $|v(b)|$ is the number of voxels of $v(b)$.  

\vspace{-0.2cm}

\subsection{Experimental design}
 
We ran multiple experiments using the multiple LAP method of \cite{sharmin2018white} (multi-LAP) and the proposed method (multi-LAP-anat) on a total of 360 segmented bundles. In both cases, each pair of tractograms were aligned with an initial affine registration. We used an example set composed of 5 bundles, since it was proved that considering a larger example set has no substantial impact on the final result of the segmentation \cite{sharmin2018white}. In the multi-LAP-anat method, we added the two anatomically-informed distance matrices to the original cost matrix as explained in section \ref{sec:anatomy_cost_matrices}.
The parameters of $\lambda_D$,  $\lambda_E$ and  $\lambda_R$ were set in order to let all the values of the matrices span in the same range (which would approximately correspond to $\lambda_D=1$, $\lambda_E=0.4$ and $\lambda_R=1.6$). To build the ROI-based distance matrix, for each bundle, we considered the two waypoint ROIs that delineate the trajectory of the bundle before it diverges towards the cortex that are defined in \cite{wakana2007reproducibility}, and we transferred them in the individual subject space through a non-linear registration. We then compared the performances of the two methods through the DSC score.



All the experiments were developed in Python code and ran using cloud computing resources provided by brainlife.io. \linebreak Code and dataset are freely available for reproducibility at https://doi.org/10.25663/brainlife.app.122 and \linebreak https://doi.org/10.25663/brainlife.pub.3 respectively.

\vspace{-0.2cm}

\subsection{Results}
\label{sec:results}

In Table \ref{tab:results} we separately report the mean DSC results for the two methods that we compared across all 30 subjects, for both small and large bundles. 
For the small bundles, with the multi-LAP method we observed a standard deviation of the mean between $0.009$ and $0.015$, and with the proposed 
multi-LAP-anat method between $0.007$ and $0.011$. For the large bundles, both methods registered a standard deviation of the mean $\leq0.005$.

\begin{table} [h!]
\small
\centering
\begin{tabular}{p{2.1cm} | p{0.6cm} | p{0.6cm} | p{0.6cm} | p{0.6cm} | p{0.6cm} | p{0.6cm}}
 & CGCl & CGCr & CGHl & CGHr & UFl & UFr \\
\hline
\hline
	 multi-LAP & 0.81 & 0.82 & 0.77 & 0.76 & 0.74 & 0.76 \\
	 \hline
	 multi-LAP-anat & \textbf{0.83} & \textbf{0.86} & \textbf{0.83} & \textbf{0.80} & \textbf{0.80} & \textbf{0.81} \\
	 \hline
\end{tabular}
\begin{tabular}{p{2.1cm} | p{0.6cm} | p{0.6cm} | p{0.6cm} | p{0.6cm} | p{0.6cm} | p{0.6cm}}
 & TRl & TRr & CSTl & CSTr & AFl & AFr \\
\hline
\hline
	 multi-LAP & 0.85 & 0.85 & 0.84 & 0.85 & 0.83 & 0.80 \\
	 \hline
	 multi-LAP-anat & \textbf{0.87} & \textbf{0.87} & \textbf{0.86} & \textbf{0.87} & \textbf{0.86} & \textbf{0.84} \\
	 \hline
\end{tabular}
\caption{\small{Mean DSC across 30 subjects for both the 6 small bundles and the 6 large bundles for the two methods compared.}}
\label{tab:results}
\end{table}

%

\begin{figure*} [!h]
\centering
\begin{minipage} [b] {0.56\textwidth}
\centering
\centering
  \includegraphics[width=4.9cm]{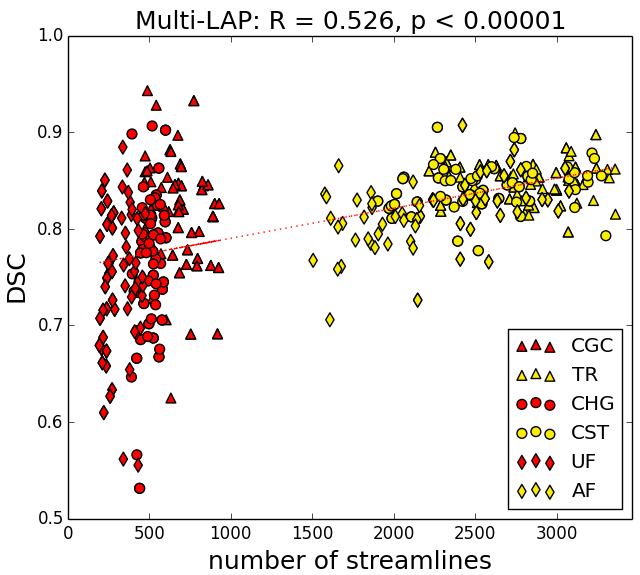} 
\hspace{0mm}
   \includegraphics[width=4.9cm]{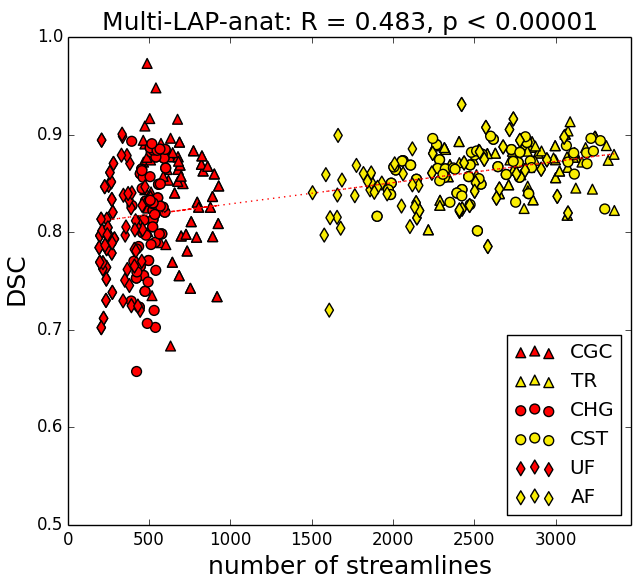}
   \caption{\small DSC as a function of bundle size using the multi-LAP method \cite{sharmin2018white} (left panel) and using the proposed multi-LAP-anat method (right panel). In red the small bundles and in yellow the large bundles, 30 examples for each bundle.}
  \label{fig:dsc-vs-str}
\end{minipage}  
\hspace{3mm}
\begin{minipage} [b] {0.39\textwidth}
\centering
   \includegraphics[width=7.2cm]{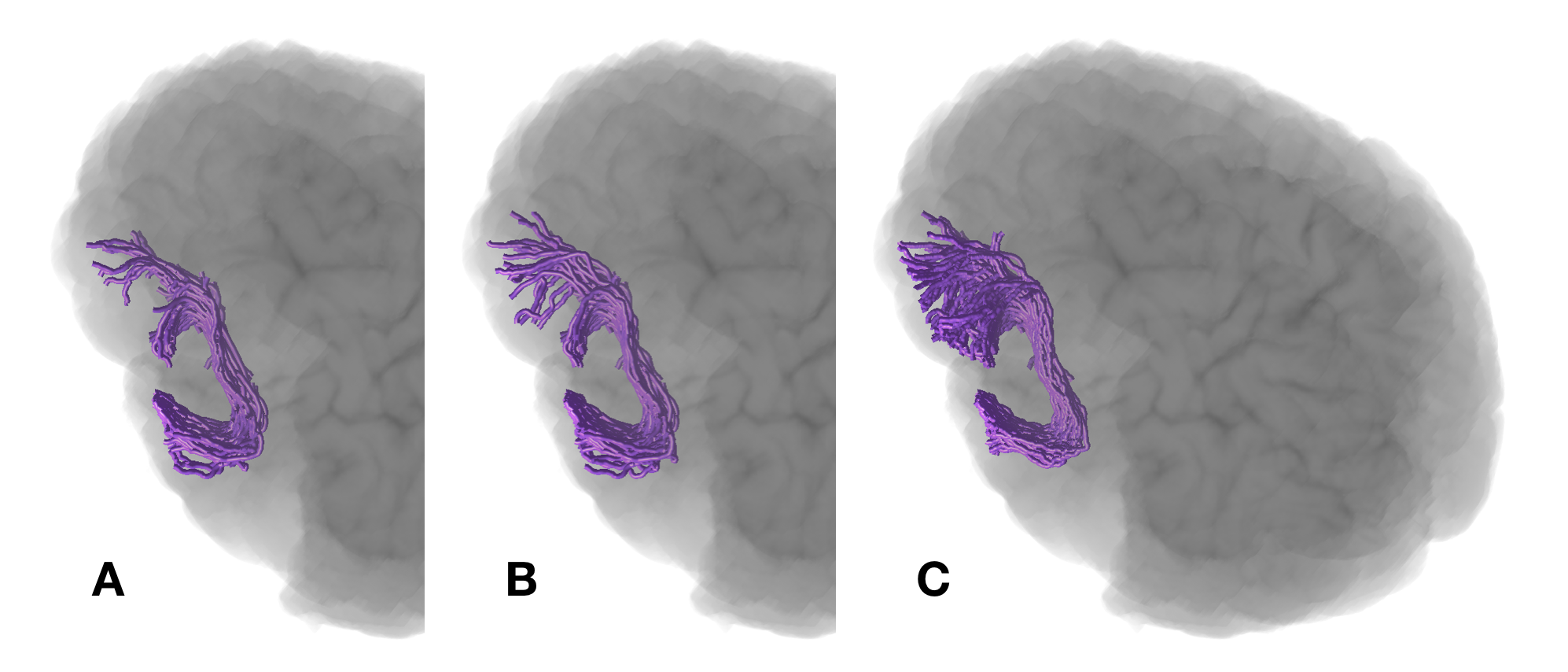}
   \caption{\small Comparative paradigmatic example of a segmented uncinate fasciculus (UF) obtained with A) the multi-LAP method \cite{sharmin2018white}, B) the proposed multi-LAP-anat method and C) the ground truth.}
  \label{fig:example}
\end{minipage}  
\end{figure*}

Figure \ref{fig:dsc-vs-str} illustrates the individual DSC scores as a function of bundle size in terms of number of streamlines, for the small and large bundles and for the two methods compared. 


%% file: discussion.tex
\section{Discussion}
\label{sec:discussion}

Table \ref{tab:results} illustrates that, on average, the proposed multi-LAP-anat method outperforms the multi-LAP method of \cite{sharmin2018white}, for all the bundles considered. In all the cases we obtained a mean DSC between 0.80 and 0.87, which means that the overlap with the ground truth is at least 80\%. 
Streamlines composing the same anatomical bundle not only have a similar shape, but are also characterized by their propensity to interconnect or pass through predefined ROIs of the brain. Including such information into the optimization process is useful in particular in identifying those streamlines that may have a less similar shape from the example, but that are close to ROIs that are known from the literature pertaining to the bundle of interest. Moreover, also taking into account the endpoint-based distance helps to select all the streamlines that end in the same terminal region. Figure \ref{fig:example} shows a paradigmatic example in which the multi-LAP-anat method correctly identifies most of the streamlines terminating in the cortical areas, which instead are partially missing in the bundle segmented by the multi-LAP method.


Table \ref{tab:results} (first row) and Figure \ref{fig:dsc-vs-str} (left panel) provide evidence that small bundles, which are usually more sensitive to registration errors, are generally harder to segment than large bundles. Using the proposed multi-LAP-anat method, we obtain a mean improvement in the DSC score of $+4.5\%$ for the small bundles, see Table \ref{tab:results}. In these cases, we also notice a decreased variance when using the proposed method, which can be seen from the comparison in Figure \ref{fig:dsc-vs-str}, where the vertical dispersion of the red points is narrower in the right panel.

The proposed method also improves the results for large bundles, for which we observe a mean improvement in the DSC score of $+2.5\%$, see Table \ref{tab:results}. These results confirm the assumption that, for all the bundles considered, including additional information about the relative anatomical position of bundles helps to improve the example-based bundle segmentation.